\documentclass[usenatbib, twocolumn]{aastex63}
\usepackage[utf8]{inputenc}
\usepackage{graphicx}
\usepackage{xspace}
\usepackage{comment}
\usepackage{savesym}
\savesymbol{tablenum}
\usepackage{siunitx}
\restoresymbol{SIX}{tablenum}
\usepackage{bm}
\usepackage{url}
\usepackage{footnote}
\usepackage{multirow}
\usepackage{booktabs}
\usepackage{verbatim}
\usepackage{textcomp}
\usepackage{lineno}
\usepackage{float}
\usepackage{tablefootnote}
\usepackage{appendix}

\begin{document}


\submitjournal{ApJ}
\shorttitle{\textsc{1991T-Like Type Ia Supernovae as an Extension of the Normal Population}}
\shortauthors{O'Brien et al.}

\title{1991T-Like Type Ia Supernovae as an Extension of the Normal Population}
\correspondingauthor{John T. O'Brien}
\email{jobrien585@gmail.com, obrie278@msu.edu}
\author[0000-0003-3615-9593]{John T. O'Brien}
\affiliation{Department of Physics and Astronomy, Michigan State University, East Lansing, MI 48824, USA}

\author[0000-0002-0479-7235]{Wolfgang E. Kerzendorf}
\affiliation{Department of Physics and Astronomy, Michigan State University, East Lansing, MI 48824, USA}
\affiliation{Department of Computational Mathematics, Science, and Engineering, Michigan State University, East Lansing, MI 48824, USA}

\author[0000-0001-7343-1678]{Andrew Fullard}
\affiliation{Department of Physics and Astronomy, Michigan State University, East Lansing, MI 48824, USA}

\author[0000-0003-3308-2420]{R\"udiger~Pakmor}
\affiliation{Max-Planck-Institut f\"ur Astrophysik, Karl-Schwarzschild-Str. 1, 85748 Garching, Germany}

\author[0000-0003-0426-6634]{Johannes Buchner}
\affiliation{Max-Planck-Institut f\"{u}r extraterrestrische Physik, Giessenbachstrasse 1, 85748 Garching bei M\"{u}nchen, Germany}

\author[0000-0002-7941-5692]{Christian Vogl}
\affiliation{Max-Planck-Institut f\"ur Astrophysik, Karl-Schwarzschild-Str. 1, 85748 Garching, Germany}
\affiliation{Exzellenzcluster ORIGINS, Boltzmannstr. 2, 85748 Garching, Germany}

\author[0000-0001-7571-0742]{Nutan Chen}
\affiliation{Machine Learning Research Lab, Volkswagen AG, Munich, Germany}

\author[0000-0003-4418-4916]{Patrick van der Smagt}
\affiliation{Machine Learning Research Lab, Volkswagen AG, Munich, Germany}
\affiliation{Faculty of Informatics, E\"otv\"os Lor\'and University, Budapest, Hungary}

\author[0000-0003-2544-4516]{Marc Williamson}
\affiliation{Department of Physics, New York University, New York, NY, 10003, USA}

\author[0000-0002-8310-0829]{Jaladh Singhal}
\affiliation{IPAC, MC 100-22, Caltech, 1200 E. California Blvd., Pasadena, CA 91125}

\begin{abstract}

 Type~Ia supernovae remain poorly understood despite decades of investigation. Massive computationally intensive hydrodynamic simulations have been developed and run to model an ever-growing number of proposed progenitor channels. Further complicating the matter, a large number of sub-types of Type~Ia supernovae have been identified in recent decades. Due to the massive computational load required, inference of the internal structure of Type~Ia supernovae ejecta directly from observations using simulations has previously been computationally intractable. However, deep-learning emulators for radiation transport simulations have alleviated such barriers. We perform abundance tomography on 40 Type~Ia supernovae from optical spectra using the radiative transfer code \textsc{TARDIS} accelerated by the probabilistic \textsc{DALEK} deep-learning emulator. We apply a parametric model of potential ejecta structures to comparatively investigate abundance distributions and internal ionization fractions of intermediate-mass elements between normal and 1991T-like Type~Ia supernovae. Our inference shows that 1991T-like Type~Ia supernovae are under-abundant in the typical intermediate mass elements that heavily contribute to the spectral line formation seen in normal Type~Ia supernovae at early times. Additionally, we find that the intermediate-mass elements present in 1991T-like Type~Ia supernovae are highly ionized compared to those in the normal Type Ia population. Finally, we conclude that the transition between normal and 1991T-like Type~Ia supernovae appears to be continuous observationally and that the observed differences come out of a combination of both abundance and ionization fractions in these supernovae populations.

\end{abstract}

\section{Introduction}

Type Ia supernovae (SNe~Ia), the thermonuclear explosions of Carbon/Oxygen (C/O) white dwarfs (WD), are critical tools for understanding the evolution of the cosmos. SNe~Ia populate galaxies with iron-group and intermediate-mass elements \citep[][see Figure 39]{Kobayashi2020} critical to the formation of planets and late-generation stars. As cosmic distance indicators \citep{Phillips1993}, SNe~Ia have proved useful in both determining the size and age of the universe, as well as for probing the nature of dark energy \citep{Branch1992Hubble, Riess1998, 1999ApJ...517..565P}. However, despite their success as tools for probing galactic and cosmological evolution, the mechanism(s) underlying their ignition remain poorly understood.

An ever-increasing number of progenitor models have been proposed in the literature to explain SNe~Ia, usually involving some sort of mass transfer from a binary companion. For example, ignition of a C/O WD has been suggested to be the result of mergers with a binary companion \citep[e.g.][]{Nomoto1982b, Webbink1984, Iben1984, van_Kerkwijk_2010, Livio_2003, Kashi2011}, accretion from a companion star onto a near Chandrasekhar-mass ($\mathrm{M_{Ch}}$) WD \citep[e.g.][]{Whelan1973} resulting in a turbulent deflagration, or accretion onto a sub-$\mathrm{M_{Ch}}$ WD resulting in a super-sonic detonation \citep[e.g.][]{Woosley1994, Fink2010, Shen2018, Polin2019,2022MNRAS.517.5260P}. Despite intensive work and an ever-increasing number of proposed models, secure progenitor identification from spectral and photometric observations remains elusive.

Further complicating the matter of progenitor identification is the large spectroscopic diversity of thermonuclear SNe that have been identified over the past few decades. A large number of objects within the class of SNe~Ia with unique spectral and photometric properties have resulted in a variety of classification schemes \citep[e.g.][]{2006PASP..118..560B, 2017hsn..book..317T}. These objects range from the subluminous low-velocity Type~Iax/02cx-like thermonuclear supernovae \citep[][]{Foley_2013} to super-luminous shallow-silicon \citep[][]{2006PASP..118..560B} 1991T-like SNe~Ia \citep[][]{1992ApJ...384L..15F,1992AJ....103.1632P}. The variation in the properties of these objects leads us to consider the possibility of either distinct progenitor channels for these sub-types or a unified progenitor model that can describe massive variations in spectral properties.

%
We begin our investigation into the relationship between SNe~Ia sub-types from the bright end of thermonuclear transients by focusing on the super-luminous 1991T-like SNe~Ia. On the observational side, 1991T-like SNe~Ia appear spectroscopically similar to the normal \citep{Branch1993, Benetti2004, Branch_2006} SNe~Ia population after their light curves achieve maximum brightness \citep{1992AJ....103.1632P}, however, in their early phases they are quite distinct. Their early-time spectra contain strong absorption lines of high-velocity Fe~II/Fe~III and lack the characteristic strong Si~II absorption features of normal SNe~Ia \citep[][]{1992ApJ...384L..15F,Filippenko1997}. Additionally, 1991T-like SNe~Ia lie close to the normal SNe~Ia in the space of the luminosity-decline rate relation, potentiality contaminating SNe~Ia samples used for cosmic distance measurements due to Malmquist bias at high redshift \citep{10.1093/mnras/stu1777}. On the theoretical side, \citet{1992ApJ...384L..15F} originally proposed that 1991T-like supernovae may either be the results of either a double-detonation initiated at an intermediate layer in the progenitor WD, or a delayed-detonation model, in order to explain the large amount of the progenitor WD that is burned into $^{56}$Ni and the apparent narrow region of IMEs present with the ejecta. Since then, many hypotheses have been proposed to explain the deviations in photometric and spectroscopic properties of 1991T-like SNe~Ia from the normal SNe~Ia population with mixed success \citep[e.g.][]{1992ApJ...387L..33R,1995A&A...297..509M, 1997ApJ...486L..35L,2015A&A...580A.118M,2016A&A...592A..57S}. A definitive connection between the theoretical progenitor channels for 1991T-like SNe~Ia and their observed spectral properties requires constraining the possible theoretical models to the observations directly. 

In this paper, we present ejecta reconstructions from inference and a direct statistical comparison of the internal ejecta state between populations of \num{35} normal and five 1991T-like SNe~Ia. The ejecta models are presented as probability distributions determined through Bayesian inference performed on single-epoch early-time optical spectra. Our parameterized ejecta model is based on hydrodynamical simulations of a variety of proposed progenitor systems from the Heidelberg Supernova Model Archive \citep[\textsc{HESMA}][]{2017MmSAI..88..312K}. We use a radiative transport scheme based on the open-source radiative transfer code \textsc{tardis} \citep{TARDIS2014} accelerated by the probabilistic \textsc{dalek} deep-learning emulator \citep{2022arXiv220909453K} to generate predictions of synthetic spectra over our space of model parameters. We compare distributions of ejecta compositions and ionization states between the normal and 1991T-like SNe~Ia populations and identify a relationship between their internal structure and observed spectral features. These results allow us to better understand the relationship between normal SNe~Ia and 1991T-like SNe~Ia.

In Section~\ref{sec:Data}, we describe the selection criteria for the observed spectra samples of normal and 1991T-like SNe~Ia that we chose to model. Section~\ref{sec:SupernovaModel} describes the parametric ejecta model implemented to model these spectra as well as details of the radiative transfer simulation and its acceleration through emulation. Section~\ref{sec:ModelInference} describes the inference framework for estimating the posterior distributions of our model parameters, including the form of the likelihood function and the priors placed on our parameters. Results of our modeling are presented in Section~\ref{sec:Results} along with a discussion of their physical implications. Finally, our conclusions and final discussion are summarized in Section~\ref{sec:Conclusion}.

\section{Data}
\label{sec:Data}

We select a sample of normal and 1991T-like SNe~Ia with spectra between \num{7} and \num{14}\, days before the B-band maximum in the light curve as these observations are well into the photospheric phase (see Section~\ref{sec:ExplosionModel}) when the ejecta are still optically thick. This selection was designed to model spectral observations taken \num{8} to \num{12}\, days post-explosion given a rise-time of \num{19.5}\, days with a \num{2.5}\, day rise-time uncertainty. Some studies \citep[e.g.][]{2022ApJ...938...47P} will discern between the transitional shallow-silicon 1999aa-like SNe~Ia and the 1991T-like SNe~Ia due to the presence of early-time Calcium features and larger Si~II absorption features. For the purposes of this study, we group together 1999aa-like SNe~Ia with 1991T-like SNe~Ia and refer to the joint group as 1991T-like SNe~Ia.

Our sample of selected SNe~Ia is based on the sample investigated by \citet{Polin_2021} as these objects are well studied. We queried \href{https://wiserep.org}{WISeREP} \citep{2012PASP..124..668Y} for each selected SN, filtering to only objects labeled as either Ia or Ia-pec with spectra within our time interval, and found a total of 158 spectra covering 44 objects. For each object found, we select a single spectrum to model according to two criteria relating to the quality and coverage of the data. We first attempt to limit our sets of spectra to those with coverage of more than 90\% of the wavelength range from $3400\,\textrm{\AA}$ to $7600\,\textrm{\AA}$ which corresponds to the wavelength range of our model. If no spectra for a single object fully encompass this range, we keep them for the next step of selection to maximize the number of objects we model. We then select the spectrum from each object with the highest average signal-to-noise ratio. If a spectrum does not include the flux error, we assume the signal-to-noise ratio for that spectrum is below that of all spectra containing a flux error column when making this cut.

We classify the spectra into two categories: 1991T-like SNe~Ia and normal SNe~Ia based on spectral template fitting. We use the Supernova Identification tool \citep[SNID][]{2007ApJ...666.1024B} to determine the sub-type, and all objects that are found to be 1991T-like objects are further investigated through a literature search (See footnotes of Table~\ref{tab:my_table}) in order to properly classify objects whose photospheric phase spectra can commonly be mistaken with 1991T-likes such as 02cx-likes/Type~Iax \citep[see e.g.][]{2022ApJ...938...47P}. The final selection includes five 1991T-like SNe~Ia and 35 normal SNe~Ia spectra. The list of objects, with their phase from maximum light, classification, and references can be found in Table~\ref{tab:my_table}. 

\begin{table*}
  \centering
  \begin{tabular}{lrlllrll}

SN & Phase (d) & $\mathrm{\lambda_{min}\,(\AA)}$ & $\mathrm{\lambda_{max}\,(\AA)}$ & Date (MJD) & Telescope & Instrument & Reference\\
\hline
\hline
\multicolumn{8}{c}{1991T-likes} \\
\hline

1991T & -9.00 & 3100.00 & 9840.00 & 48365.00 & Lick-3m & UV-Schmidt  &  \citet{1992ApJ...384L..15F} \\
2001V\footnote{\citet{Zheng_2018} reports this object as a normal SNe~Ia but our results from SNID classify this as a 1991T-like SNe~Ia which we keep based on the high-brightness and low Si~II velocity.} & -9.67 & 3720.00 & 7540.50 & 51963.33 & FLWO-1.5m & FAST  &  \citet{2008AJ....135.1598M} \\
2003fa\footnote{\label{note:99aa}\citet{Zheng_2018} classifies these as a 1999aa-like SNe~Ia} & -9.66 & 3720.00 & 7540.50 & 52797.34 & FLWO-1.5m & FAST  &  \citet{2012AJ....143..126B} \\
1999dq$^{\mathrm{\ref{note:99aa}}}$ & -9.55 & 3380.00 & 9040.00 & 51426.45 & FLWO-1.5m & FAST  &  \citet{2008AJ....135.1598M} \\
1999aa & -11.67 & 3440.00 & 7220.00 & 51223.33 & FLWO-1.5m & FAST  &  \citet{2008AJ....135.1598M} \\
\hline
\multicolumn{8}{c}{Normal} \\
\hline
1998dm & -11.49 & 3300.00 & 10100.00 & 51049.51 & Lick-3m & KAST  &  \citet{2012MNRAS.425.1789S} \\
2005ki & -8.50 & 3708.77 & 7151.80 & 53697.00 & LCO-duPont & Mod-spec  &  \citet{sne.space} \\
2005mz & -7.67 & 3490.00 & 7409.02 & 53738.13 & FLWO-1.5m & FAST  &  \citet{2012AJ....143..126B} \\
2006X & -10.00 & 4134.97 & 6794.63 & 53775.00 & Nayuta & MALLS  &  \citet{2009PASJ...61..713Y} \\
2006ax & -8.70 & 3486.00 & 7407.96 & 53818.30 & FLWO-1.5m & FAST  &  \citet{2012AJ....143..126B} \\
2006cp & -9.74 & 3482.00 & 7403.96 & 53887.26 & FLWO-1.5m & FAST  &  \citet{2012AJ....143..126B} \\
2006gr & -7.70 & 3479.00 & 7415.66 & 54005.30 & FLWO-1.5m & FAST  &  \citet{2012AJ....143..126B} \\
2000dn & -7.91 & 3720.00 & 7540.50 & 51816.29 & FLWO-1.5m & FAST  &  \citet{2012AJ....143..126B} \\
2006lf & -7.60 & 3477.00 & 7413.66 & 54037.40 & FLWO-1.5m & FAST  &  \citet{2012AJ....143..126B} \\
2007af & -10.00 & 3182.61 & 5271.20 & 54163.00 & ESO-NTT & EMMI  &  \citet{sne.space} \\
2007bd & -9.32 & 3476.00 & 7412.66 & 54197.18 & FLWO-1.5m & FAST  &  \citet{2012AJ....143..126B} \\
2007ci & -8.20 & 3480.00 & 7416.66 & 54238.20 & FLWO-1.5m & FAST  &  \citet{2012AJ....143..126B} \\
1998dh & -8.50 & 3720.00 & 7540.50 & 51021.40 & FLWO-1.5m & FAST  &  \citet{2008AJ....135.1598M} \\
2007qe & -8.89 & 3476.00 & 7417.07 & 54420.11 & FLWO-1.5m & FAST  &  \citet{2012AJ....143..126B} \\
1998aq & -7.74 & 3499.50 & 7140.00 & 50922.26 & FLWO-1.5m & FAST  &  \citet{2003AJ....126.1489B} \\
2005cf & -8.71 & 3485.00 & 7411.37 & 53524.29 & FLWO-1.5m & FAST  &  \citet{2009ApJ...697..380W} \\
2006le & -7.57 & 3476.00 & 7412.66 & 54040.43 & FLWO-1.5m & FAST  &  \citet{2012AJ....143..126B} \\
2004eo & -10.00 & 3741.26 & 9092.24 & 53268.00 & LCO-duPont & WFCCD  &  \citet{sne.space} \\
2004at & -7.58 & 3720.00 & 7540.50 & 53084.42 & FLWO-1.5m & FAST  &  \citet{2012AJ....143..126B} \\
2000fa & -11.52 & 3680.00 & 7541.00 & 51881.48 & FLWO-1.5m & FAST  &  \citet{2008AJ....135.1598M} \\
2001ep & -7.51 & 3720.00 & 7540.50 & 52192.49 & FLWO-1.5m & FAST  &  \citet{2012AJ....143..126B} \\
2001gc & -8.64 & 3720.00 & 7540.50 & 52235.26 & FLWO-1.5m & FAST  &  \citet{2012AJ....143..126B} \\
2002bo & -7.66 & 3720.00 & 7540.50 & 52349.34 & FLWO-1.5m & FAST  &  \citet{2012AJ....143..126B} \\
2002cr & -11.31 & 3720.00 & 7540.50 & 52397.29 & FLWO-1.5m & FAST  &  \citet{2012AJ....143..126B} \\
2002cs & -8.61 & 3720.00 & 7540.50 & 52401.39 & FLWO-1.5m & FAST  &  \citet{2012AJ....143..126B} \\
2004ef & -8.70 & 3479.00 & 7414.19 & 53255.30 & FLWO-1.5m & FAST  &  \citet{2012AJ....143..126B} \\
2002dj & -7.83 & 3720.00 & 7560.00 & 52443.17 & FLWO-1.5m & FAST  &  \citet{2012AJ....143..126B} \\
2002er & -8.00 & 3500.47 & 9294.97 & 52516.00 & Ekar & AFOSC  &  \citet{2005AA...436.1021K} \\
2002he & -8.52 & 3720.00 & 7500.00 & 52577.48 & FLWO-1.5m & FAST  &  \citet{2012AJ....143..126B} \\
2003W & -11.65 & 3200.00 & 8800.00 & 52668.35 & MMT & MMT-Blue  &  \citet{2012AJ....143..126B} \\
2003cg & -8.00 & 3700.00 & 9347.83 & 52721.00 & CA-2.2m & CAFOS  &  \citet{2006MNRAS.369.1880E} \\
2003du & -7.76 & 3720.00 & 7540.50 & 52757.24 & FLWO-1.5m & FAST  &  \citet{2012AJ....143..126B} \\
2008ar & -8.71 & 3476.00 & 7418.54 & 54525.39 & FLWO-1.5m & FAST  &  \citet{2012AJ....143..126B} \\
2002dl & -7.55 & 3720.00 & 7540.50 & 52444.45 & FLWO-1.5m & FAST  &  \citet{2012AJ....143..126B} \\
2011fe & -11.00 & 3500.91 & 9498.69 & 55803.00 & WHT-4.2m & ISIS  &  \citet{2012ApJ...752L..26P} \\

\end{tabular}
  \caption{Table of selected SNe with photospheric phase spectra. The phase of the spectrum represents the time before maximum B-band magnitude that the spectrum was taken. Classification of the SNe~Ia sub-types was performed with SNID for all models and further classification of those initially labeled as 91T-likes is determined through a literature search to avoid possible contamination.}
  \label{tab:my_table}
\end{table*}


\section{Supernova Model}
\label{sec:SupernovaModel}
We present a condensed parametric ejecta model designed to fit a wide variety of predicted SNe~Ia spectra corresponding to different progenitor systems. In Section~\ref{sec:ParameterizedEjectaModel} we introduce the hydrodynamic models upon which these parameters and their ranges are based. Section~\ref{sec:DensityProfile} introduces the way that the density structure of the ejecta is parameterized in the regime of the photospheric outer ejecta. Section~\ref{sec:AbundanceProfile} describes the method by which we parameterize the relative abundances according to the masses of individual elements present throughout the ejecta and how these masses are folded into a general multi-zone model for SNe~Ia ejecta. Sections~\ref{sec:ExplosionModel} and \ref{sec:RadiativeTransfer} describe the physical assumptions made when performing spectral synthesis for comparison between model parameters and observed spectra. Finally, Section~\ref{sec:Emulator} describes the deep-learning framework implemented to perform the acceleration of our spectral synthesis over our space of model parameters.

\subsection{Parameterized Ejecta Model}
\label{sec:ParameterizedEjectaModel}
We develop a parametric model of the ejecta of SNe~Ia based on the structure of spherically averaged ejecta profiles taken from \textsc{HESMA}. HESMA contains a database of a wide range of simulations of a variety of proposed SNe~Ia progenitor scenarios \citep{2014MNRAS.438.1762F,2017MNRAS.472.2787N,2013MNRAS.429.2287K,2015MNRAS.450.3045K,2010ApJ...714L..52S,2017MNRAS.472.2787N,2018A&A...618A.124F,2015A&A...580A.118M,2010A&A...514A..53F,2010ApJ...719.1067K,2012MNRAS.420.3003S,2020A&A...635A.169G} which provide an approximation to the space of potential ejecta structures that describe SNe~Ia observations at various times. A visualization of a randomly generated ejecta profile from a set of model parameters drawn from our space is presented in Figure~\ref{fig:eprofile}. The ejecta model is parameterized by density and abundance profiles, described in the next two sections.

\subsubsection{Density Profile}
\label{sec:DensityProfile}
We adopt a velocity-dependent power-law density profile in homologous expansion to model the outer ejecta of the supernova (Equation~\ref{eq:density}). The outer ejecta of HESMA models can be well fit by power-law at early times. A power-law index, $\alpha_{\rho}$ is left as a free parameter which allows the model to cover the full range of outer-ejecta density profiles present in the HESMA models (see Section~\ref{sec:PriorBounds} for a description). A reference velocity for our density profile, $v_{0}=\mathrm{8000\ km\,s^{-1}}$, is statically set for all models as a reference density, $\rho_{0}$, is solved to constrain the density of the model. 
The constructed density profile extends from $v_{0}$ to an outer boundary velocity, $v_\mathrm{outer}$, set such that the density at the outer-boundary velocity is $\rho(v_\mathrm{outer}, t=t_{0})=10^{-14}\ \mathrm{g\,cm^{3}}$ which is the cutoff value of the density profiles present in the HESMA models at $t_{0}=2\,\mathrm{days}$.
The value of $v_{0}$ is an arbitrary choice as a reference coordinate from where we define our model, so the value was selected as the lower bound of the inner boundary velocity prior (Section~\ref{sec:PriorBounds}) for simplicity.

\begin{equation} \label{eq:density}
\rho(v) = \rho_{0}\left(\frac{t_{0}}{t}\right)^{3}\left(\frac{v}{v_{0}}\right)^{\alpha_{\rho}}
\end{equation}

We constrain the values for $v_\mathrm{outer}$ and $\rho_{0}$ from a given total ejecta mass above $v_{0}$, $\mathrm{M_\mathrm{tot}}$, and a given $\alpha_{\rho}$ by integrating Equation~\ref{eq:density} at a time $t=t_{0}$ by applying the substitution $v\,t_{0}=r$ from homologous expansion.

\begin{equation} \label{eq:mass}
\mathrm{M_{tot}} = \frac{\rho_{0} t_{0}^3 4\pi}{v_{0}^{\alpha_{\rho}}}\int_{v_{0}}^{v_\mathrm{outer}} v^{\alpha_{\rho}} v^{2}dv
\end{equation}

The value for $\mathrm{M_{tot}}$ is determined from the total of the masses of the individual elements contributing to the ejecta above $v_{0}$. 

\subsubsection{Abundance Profile}
\label{sec:AbundanceProfile}

\begin{figure}
  \centering
  \includegraphics[width=0.49\textwidth]{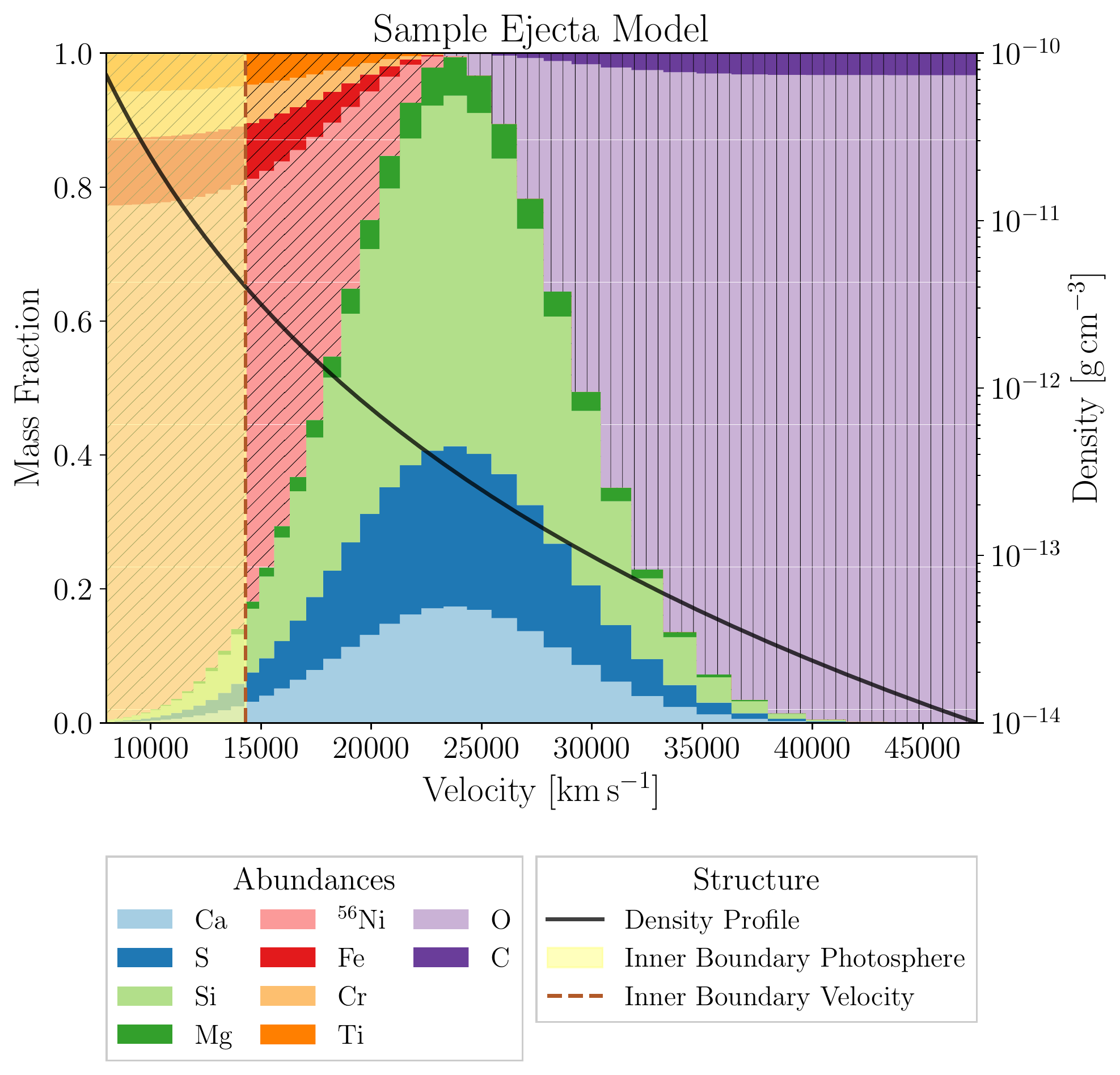}
  \caption{Visualization of a random realization of an abundance profile produced from our model in our prior space. Elemental abundances are presented as stacked histograms. The diagonally hatched regions correspond to the inner iron-group elements, the central unhatched region corresponds to the abundance of IMEs, and the vertically hatched region corresponds to the unburned elements in the outer layers of the ejecta. A red dashed vertical line represents the inner boundary velocity from which thermalized radiative packets are injected into the ejecta above. The solid black line represents the density of the ejecta through velocity space and the value of the density if provided by the right-hand axis.}
  \label{fig:eprofile}
\end{figure}

We model the abundances of the same elements explored by \citet{2021ApJ...916L..14O} in our ejecta model as these elements account for the majority of line formation in the resulting spectrum as well as trace the general nucleosynthetic products of the supernova \citep[see e.g.][]{Filippenko1997}.
We parameterize these elements in terms of total masses above $v_{0}$ in order to better constrain the total ejecta mass as well as simplify the sampling procedure.
Masses for Carbon ($\mathrm{M_{C}}$), Oxygen ($\mathrm{M_{O}}$), Magnesium ($\mathrm{M_{Mg}}$), Silicon ($\mathrm{M_{Si}}$), Sulfur ($\mathrm{M_{S}}$), Calcium ($\mathrm{M_{Ca}}$), Chromium ($\mathrm{M_{Cr}}$), Titanium ($\mathrm{M_{Ti}}$), stable Iron ($\mathrm{M_{Fe}}$), and initial $^{56}$Ni at $t_{0}$, $\mathrm{M_{^{56}Ni}}$, are aggregated into three quantities corresponding to the mass of Iron Group Elements (IGEs, $\mathrm{M_{IGE}} = \mathrm{M_{^{56}Ni} + M_{Cr} + M_{Ti} + M_{Fe}}$), Intermediate Mass Elements (IMEs, $\mathrm{M_{IME}} = \mathrm{M_{Si} + M_{S} + M_{Mg} + M_{Ca}}$), and Unburned Elements (UBEs, $\mathrm{M_{UBE}} = \mathrm{M_{C} + M_{O}}$), as well as a total ejecta mass ($\mathrm{M_{tot}} = \mathrm{M_{IGE} + M_{IME} + M_{UBE}}$). We place these three categories of elements into three distinct regions of the ejecta corresponding to a general structure seen in the HESMA abundance profiles as well as tomography results presented by \citet[][Figure 18]{2022MNRAS.515.4445A} in which IGEs resulting from complete nuclear burning are placed below a layer of IMEs resulting from incomplete burning, with UBEs placed in the outer-most regions (see Figure~\ref{fig:eprofile}). The fractional abundance of each region is parameterized by a set of functions, $A_{UBE}(v;v_{c},w)$, $A_{IME}(v;v_{c},w)$, $A_{IGE}(v;v_{c},w)$, where the sum of the profiles at each velocity adds up to unity. A modified Gaussian is used to represent the distribution of IMEs which is parameterized by a width, $w$, and a centroid, $v_{c}$, in velocity space. The form of this profile was selected to allow for the model to parameterize various amounts of mixing between regions of the ejecta as well as explore the depth at which the properties of the ejecta are changing. The model results in a mass-fraction profile that follows a Gaussian bubble of IMEs over the ejecta velocity and serves as an approximation to the profiles present in the HESMA dataset.

\begin{equation} \label{eq:AIME}
A_\mathrm{IME}(v;v_{c},w) = A_{0} v^{-(\alpha_{\rho}+2)} \exp\left[-\frac{1}{2}\frac{(v - v_{c})^{2}}{w^{2}}\right]
\end{equation}
Where $A_{0}$ is a normalization constant set to the inverse of the maximum value of $A_{IME}(v=v_\mathrm{max};v_{c},w)$. The velocity corresponding to the distribution's maximum value is determined from $v_{c}$ and $w$ through the relation
\begin{equation} \label{eq:vc}
v_{c} = \frac{w^{2} (\alpha_{\rho} + 2) + v_\mathrm{max}}{v_\mathrm{max}}.
\end{equation}

The values for $v_\mathrm{max}$ and $w$ are then determined from the relative masses of each region of elements by numerically solving the following system of equations
\begin{equation} \label{eq:MIME}
\mathrm{M_{IME}} = \frac{\rho_{0} t_{0}^3 4\pi}{v_{0}^{\alpha_{\rho}}}\int_{v_{0}}^{v_\mathrm{outer}} v^{\alpha_{\rho}} A_\mathrm{IME}(v) v^{2}dv,
\end{equation}
\begin{equation} \label{eq:MIGE}
\mathrm{M_{IGE}} = \frac{\rho_{0} t_{0}^3 4\pi}{v_{0}^{\alpha_{\rho}}}\int_{v_{0}}^{v_\mathrm{max}}v^{\alpha_{\rho}}\left[1-A_\mathrm{IME}(v)\right] v^{2}dv,
\end{equation}
\begin{equation} \label{eq:MUBE}
\mathrm{M_{UBE}} = \frac{\rho_{0} t_{0}^3 4\pi}{v_{0}^{\alpha_{\rho}}}\int_{v_\mathrm{max}}^{v_\mathrm{outer}} v^{\alpha_{\rho}} \left[1-A_\mathrm{IME}(v)\right] v^{2} dv
\end{equation}
which results in a complete ejecta profile. 

\subsubsection{Explosion Model}
\label{sec:ExplosionModel}

Our analysis of early-phase spectra relies on the photospheric approximation in which the rapidly increasing optical depth of the ejecta towards the center is approximated as a hard inner boundary in velocity space, $v_{inner}$. Thermalized radiation is injected into the ejecta above from a black-body distribution at a given temperature, $T_{inner}$. A parameter representing the time since the explosion, $t_{exp}$, scales the density profile (Equation~\ref{eq:density}) as well as sets the abundances of decay products of $\mathrm{^{56}Ni}$ in the final ejecta profile.

\subsection{Radiative Transfer\label{sec:RadiativeTransfer}}

We calculate synthetic spectra from our ejecta model using the open-source Monte Carlo radiative transfer code \textsc{tardis} \citep{TARDIS2014,kerzendorf_wolfgang_2021_4995779}. \textsc{Tardis} is a 1D steady-state code that iteratively solves for the excitation and ionization state of the plasma. \textsc{tardis} uses an inner boundary photosphere approximation that injects radiative packets into a homologously expanding ejecta. 

In this work, we use \textsc{tardis} version \texttt{022.5.9.dev5+gf27fa30} together with atomic data being produced by the \textsc{tardis} sub-package \textsc{carsus} \citep{passaro_e_a_2019_4062427} version \texttt{0.1.dev677+gd623c94}. The generated atomic data takes ionization energies from \textsc{CMFGEN} \citep{2001ASPC..247..343H} for O~I, O~II, S~I, S~II, Si~I, and Si~II. Ionization energies for other species used in this work were taken from \textsc{NIST} \citep{2005MSAIS...8...96R} with lines and level data taken from Kurucz \textsc{GFALL} \citep{1995KurCD..23.....K}.

Ionization populations are solved using the ``nebular'' approximation \citep[Equation 3 in][]{TARDIS2014} and excitation populations are solved using the ``dilute-lte'' prescription (Equation 5 in \citeauthor{TARDIS2014} \citeyear{TARDIS2014}; Equation 4 in \citeauthor{1999AA...345..211L} \citeyear{1999AA...345..211L}). Line interactions are handled using a macro-atom model \citep{2002AA...384..725L}. Models were generated using 40 shells of ejecta and run until plasma state convergence with $10^{5}$ packets per Monte Carlo iteration. Further configuration information for \textsc{tardis} including links to a reproducible setup and the atomic data file created with \textsc{carsus} can be found in Appendix~\ref{Appendix:A}.

\subsubsection{Emulator}
\label{sec:Emulator}


Spectral synthesis with \textsc{tardis} is too computationally expensive to be used directly for fitting. For example, a single \textsc{tardis} simulation takes approximately 30 CPU minutes, which would require hundreds of years to effectively sample a posterior distribution which requires over a million sequential simulation runs. 

In recent years emulation of radiative transfer models has served as a powerful tool for directly probing the properties of a variety of supernovae and other astrophysical objects \citep[see e.g.][]{2020A&A...633A..88V, 2021ApJ...916L..14O, Fullard_2022}. To expedite model evaluation we employ an emulator for \textsc{tardis} which performs spectral synthesis from model input parameters through an analytic approximation. \citet{2021ApJ...916L..14O} applied a deep-learning emulator for \textsc{tardis} based on \citet{Dalek2020} to simulate single-zone ejecta models for normal SNe~Ia which, for the first time, allowed for fully-probabilistic reconstructions of the outer ejecta of a SN~Ia. \citet{2022arXiv220909453K} expanded upon the utility of such emulators by incorporating a probabilistic deep-learning architecture for emulated SNe~Ia spectral synthesis which includes the added functionality of providing uncertainties in the emulated spectra.

We combine our ejecta model with the probabilistic emulator architecture to rapidly generate synthetic spectra from our model's parameters with improved uncertainty estimates. We train a deep ensemble \citep[][]{lakshminarayanan2017simple} of 12 probabilistic emulators to emulate our spectral synthesis. Model evaluation is performed by aggregating the resulting spectra from each emulator with their associated uncertainty. Scripts and data files containing the emulator and its training data can be found in Appendix~\ref{Appendix:A}.

\section{Model Inference}
\label{sec:ModelInference}

We perform Bayesian inference in order to find the posterior distribution of model parameters given our observed spectra. In order to model the posterior distribution we require a method of likelihood estimation, presented in Section~\ref{sec:LikelihoodEstimation}, to effectively compare simulated spectra to observed spectra in the context of physical and systematic uncertainties and biases. The constraints we place on the parameters of our model are discussed in Section~\ref{sec:PriorBounds} and the method of sampling the posterior distribution is discussed in Section~\ref{sec:SamplingThePosterior}. A short discussion of our method of lowering the emulation uncertainty for regions of parameter space that are both high in likelihood and under-sampled in our emulator's training data is presented in Section~\ref{sec:ActiveLearning}.

\subsection{Likelihood Estimation}
\label{sec:LikelihoodEstimation}



We apply an extended form of the likelihood function used by \citet{2021ApJ...916L..14O} to incorporate emulator uncertainties determined by the probabilistic \textsc{Dalek} emulator by adding them in quadrature to the other sources of uncertainty. We aim to best reconstruct the composition of the ejecta, so we remove the continuum when determining the quality of a fit in order to maximize contributions from line formation. We incorporate a spectral continuum removal process, $C(\hat{F}_{\lambda}(\vec{\theta}))$ which normalizes the synthetic spectrum estimate, $\hat{F}_{\lambda}(\vec{\theta})$ to the continuum of the observed spectrum, $F_{\lambda}$. This continuum removal process fits a 3rd order polynomial to the ratio between the observed spectrum and the simulated spectrum then multiplies the simulated spectrum by the polynomial. Such removal is necessary to remove the effects of the continuum, distance, and reddening from the observation to ensure our fits are driven by the line features. The total form of the log-likelihood is

$$\log \mathcal{L}(\vec{\theta}) = -\frac{1}{2}\sum_{\lambda} \left[\left(\frac{C(\hat{F}_{\lambda}(\vec{\theta})) - F_{\lambda}}{\sigma_{\lambda}(\vec{\theta})}\right)^{2} + \log{\left(2\pi\sigma_{\lambda}^{2}(\vec{\theta})\right)}\right],$$
where
$$\sigma^{2}_{\lambda}(\vec{\theta}) = \sigma_{\mathrm{obs}, \lambda}^{2} + f_{\sigma}^{2}C^{2}(\hat{F}_{\lambda}(\vec{\theta})) + \sigma_{\mathrm{emu}, \lambda}^{2}(\vec{\theta})$$
where $f_{\sigma}$ represents an inferred fractional uncertainty \citep{Hogg2010} over our spectrum and $\sigma_{\mathrm{obs}, \lambda}$ is the observational uncertainty of the spectrum we are fitting. Observational uncertainties are taken from the spectra data source if available, otherwise, a constant uncertainty of 1\% of the mean of the spectrum is assumed. $\sigma_{\mathrm{emu}, \lambda}$ is the estimate of the emulator's uncertainty \citep[Equation~\num{4} in][]{2022arXiv220909453K} in the region corresponding to the fit.

\subsection{Prior Bounds}
\label{sec:PriorBounds}

\begin{table}[ht]
  \centering
  
\begin{tabular}{c|c|c|c}

Distribution & Model Parameter & \multicolumn{2}{c}{Distribution Parameters} \\
\hline
Uniform & & Low & High  \\

 & $T_\mathrm{inner}\ (\mathrm{K})$\footnote{Prior distributions for $T_\mathrm{inner}$, $v_\mathrm{inner}$ and $t_\mathrm{exp}$ are further constrained by the condition that the luminosity estimated from the Stephan-Boltzmann law $8\times10^{40}\ \mathrm{erg\,s^{-1}} < 4\pi\sigma_{SB} v_\mathrm{inner}^{2}t_\mathrm{rise}^{2}T_\mathrm{inner}^{4} < 5\times10^{43}\ \mathrm{erg\,s^{-1}}$ based on the estimated range of SNe~Ia luminosities computed from Figure~1 of \citet{2017hsn..book..317T} \label{note:lum}} & 8000 & 15000 \\
& $v_\mathrm{inner}\ (\mathrm{km\,s^{-1}})$$^{\mathrm{\ref{note:lum}}}$ & 8000 & 16000 \\
& $\alpha_{\rho}$ & -10 & -5 \\
\hline
Normal & & $\mu$\footnote{Mean of the normal distribution in linear space.} & $\sigma$\footnote{Standard deviation of the normal distribution} \\
 & $t_\mathrm{exp}\, (\mathrm{days})$ & \num{19.5} + Phase\footnote{Prior centroid is dependent on the phase of the spectrum from maximum light reported in Table~\ref{tab:my_table}} & $2.5$ \\
\hline
Multivariate  & & $\mu$\footnote{Mean of the prior distribution in linear space.  The centroid of the log-normal distribution is the $\log_{10}$ of this values.} & $\sigma$\footnote{1D standard deviation of the $\log_{10}$ of each mass distribution.  It is important to note that there exists a non-zero covariance between each mass term.} \\
Log-normal & $\mathrm{M_{Si}}\, (\mathrm{M_{\odot}})$ & $7.84\times10^{-2}$ & 0.93 \\
& $\mathrm{M_{Ca}}\, (\mathrm{M_{\odot}})$ & $1.10\times10^{-2}$ & 1.08 \\
& $\mathrm{M_{S}}\, (\mathrm{M_{\odot}})$ & $3.94\times10^{-2}$ & 0.89 \\
& $\mathrm{M_{Mg}}\, (\mathrm{M_{\odot}})$ & $1.19\times10^{-2}$ & 1.02 \\
& $\mathrm{M_{Ni56}}\, (\mathrm{M_{\odot}})$ & $1.11\times10^{-1}$ & 1.49 \\
& $\mathrm{M_{Cr}}\, (\mathrm{M_{\odot}})$ & $3.17\times10^{-3}$ & 1.47 \\
& $\mathrm{M_{Ti}}\, (\mathrm{M_{\odot}})$ & $1.48\times10^{-3}$ & 1.87 \\
& $\mathrm{M_{Fe}}\, (\mathrm{M_{\odot}})$ & $2.04\times10^{-2}$ & 1.39 \\
& $\mathrm{M_{O}}\, (\mathrm{M_{\odot}})$ & $7.12\times10^{-2}$ & 1.34 \\
& $\mathrm{M_{C}}\, (\mathrm{M_{\odot}})$ & $2.59\times10^{-2}$ & 0.87 \\

\end{tabular}

  \caption{The prior distributions from which our model parameters are sampled during posterior inference. Parameters are sampled over different distributions according to their range of physical applicability determined from hydrodynamical models in the HESMA data set.}
  \label{tab:priors}
\end{table}



Table~\ref{tab:priors} lists our prior distributions of model parameters.
Multiple constraints are placed on the prior distribution of model parameters in order to accurately reflect the limits of currently explored hydrodynamic simulations of progenitor scenarios for SNe~Ia.
A large variety of hydrodynamical simulations of various SNe~Ia progenitor systems are found in the HESMA models and offer information about the expected general properties of the ejecta structure such as the relative typical ratios of nucleosynthetic products present within the ejecta as well as full density profiles. We generate a prior space for total elemental masses by integrating models taken from HESMA above $v_{0}$ so that the final masses of each element follow the same general correlation structure as the sum of all hydrodynamic models, ensuring a reasonable estimate of the distribution of likely supernovae ejecta profiles. The prior distribution of elemental masses is drawn from a multivariate Gaussian distribution whose covariance is set as the covariance of the log of elemental masses taken from the HESMA models with a centroid taken as the log of the mean of HESMA masses in linear space as to not bias the distribution towards models with little or no mass of certain elements. Drawing from this distribution offers a good balance between tracing the general covariance structure of the models found in the HESMA while also permitting nearly any parameter combination to be tested, albeit with a smaller probability. 

We set a uniform prior on the distribution of values of $\alpha_{\rho}$ by fitting linear models to the HESMA density profiles above $v_{0}$ and taking the minimum and maximum value to the nearest integer.
Velocity and temperature distributions are initially sampled uniformly over the ranges specified in Table~\ref{tab:priors}, with cuts placed on the luminosity of the supernovae under homologous expansion with an assumed rise time of \num{19.5}\, days \citep{Riess_1999b} according to the Stephan-Boltzmann law as an estimate for the range of realistic maximum light luminosities.
The prior distribution for the time since the explosion, $t_{exp}$, is determined on a spectrum-by-spectrum basis. The distribution is always represented by a Gaussian distribution centered at a time of \num{19.5}\,days plus the phase of the spectrum from maximum light (see Table~\ref{tab:my_table}) with a standard deviation of \num{2.5}\,days to account for rise-time uncertainty based on the spread of rise-times between normal and 1991T-like SNe~Ia \citep[see Figure~6 in][]{10.1111/j.1365-2966.2011.19213.x}. 

\subsection{Sampling the posterior}
\label{sec:SamplingThePosterior}
\subsubsection{UltraNest}
\label{sec:UltraNest}

The posterior inference was performed with nested sampling \citep{Skilling2004, Buchner2021} with the MLFriends Monte Carlo algorithm \citep{Buchner2014, Buchner2017}. Nested sampling is ideal for generating posterior samples from complex high-dimensional distributions. We used the nested sampling package \textsc{UltraNest}\footnote{\url{https://johannesbuchner.github.io/UltraNest/}} \citep{BuchnerJoss2021} to sample the posterior distribution for each observed spectrum. Each spectrum returned between \num{10000} and \num{30000} effective posterior samples which are presented in Figures~\ref{fig:EjectaStructure} and~\ref{fig:ElemMassFraction}.

\subsubsection{Active Learning}
\label{sec:ActiveLearning}

The high dimensionality of the parameter space and unknown apriori parameter constraints required to effectively model individual spectra observations create difficulty in selecting an optimal training set for our emulator. We resolve this issue by iteratively selecting new training points that are predicted to best improve emulator accuracy in the regions of the parameter space that are most likely to model the spectra we are attempting to model.

We apply \textit{Active Learning} \citep[AL][]{cohn1996active, beluch2018power} iterations to the emulator training to improve accuracy in regions of high importance. After an initial draw of \num{250000} random samples, the emulator is trained to reproduce the results of \textsc{tardis} (see Section~\ref{sec:Emulator}). We sample the posterior distribution, using this emulator, of parameters best matching our observed spectra using a modified AL likelihood function, $\mathcal{L}_\mathrm{{AL}}(\vec{\theta})$. This likelihood function weighs the likelihood of a proposed $\vec{\theta}$ by the relative fraction of emulator uncertainty to total uncertainty, encouraging exploration into regions of the parameter space where the emulator has less information.
The AL likelihood function is computed as
$$\log \mathcal{L}_\mathrm{{AL}}(\vec{\theta}) = \log \mathcal{L}(\vec{\theta}) + \frac{1}{2}\sum_{\lambda} \log{\frac{\sigma_{\mathrm{emu}, \lambda}^{2}(\vec{\theta})}{\sigma^{2}_{\lambda}(\vec{\theta})}}$$

An equal number of posterior samples are selected for each observed spectrum and are evaluated by \textsc{tardis}. Synthetic \textsc{tardis} spectra are then appended to the original training data to provide the emulator with more information around areas that are simultaneously high in likelihood while also high in emulation uncertainty. Each acquisition process yields approximately \num{200000} additional samples per iteration. Two iterations of active learning were performed on the data.

\section{Results}
\label{sec:Results}


\begin{figure}
  \centering
  \includegraphics[width=0.5\textwidth]{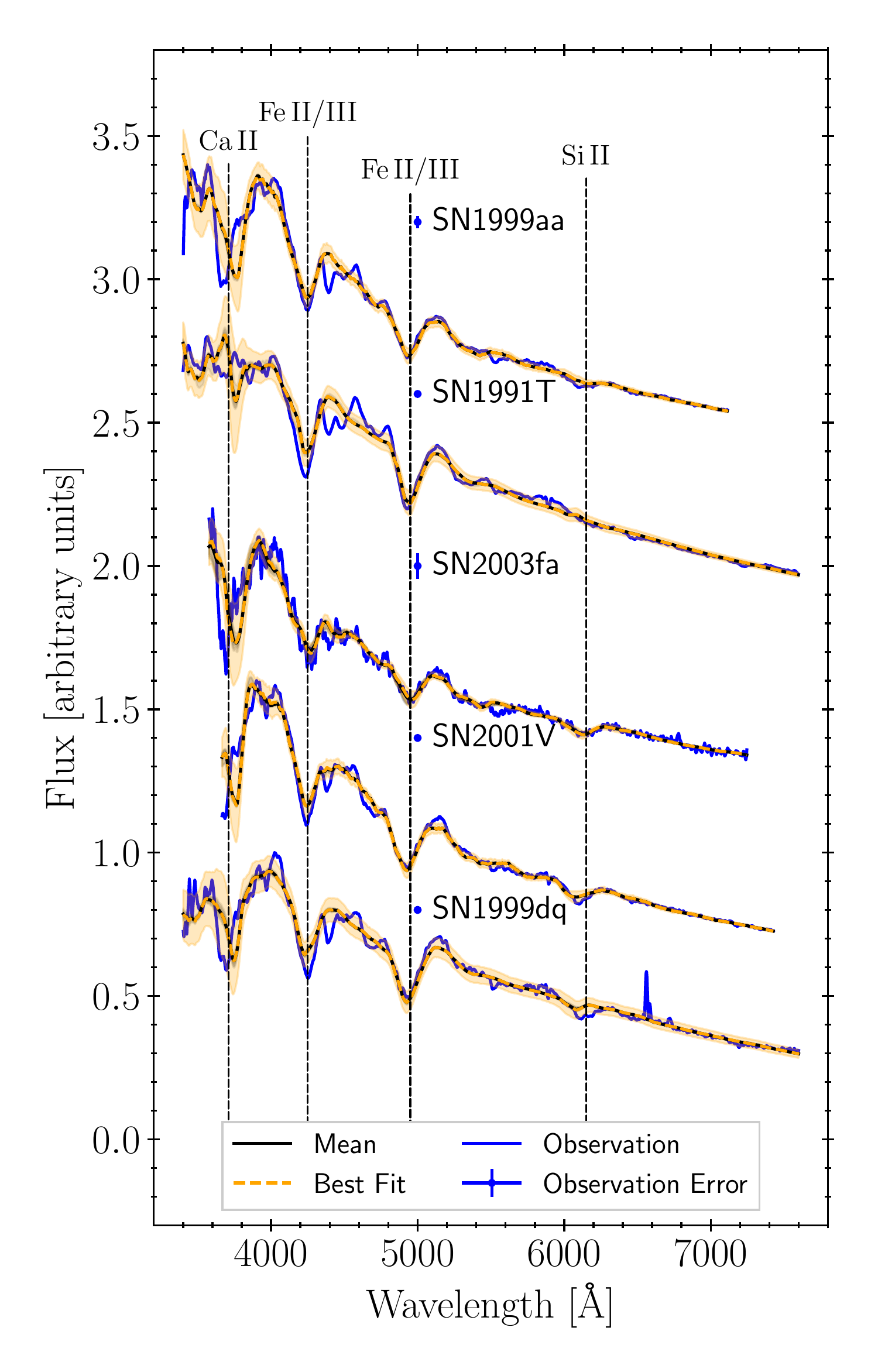} 
  
  \caption{Posterior spectra of 1991T-like SNe~Ia scaled and offset for visualization. The mean of the posterior is represented in black with the best fit (maximum likelihood sample) in orange dashed and the shaded orange region representing the total uncertainty of the best-fit sample at 1-$\sigma$.}
  \label{fig:Spectra}
\end{figure}

\begin{figure}
  \centering
  \includegraphics[width=0.5\textwidth]{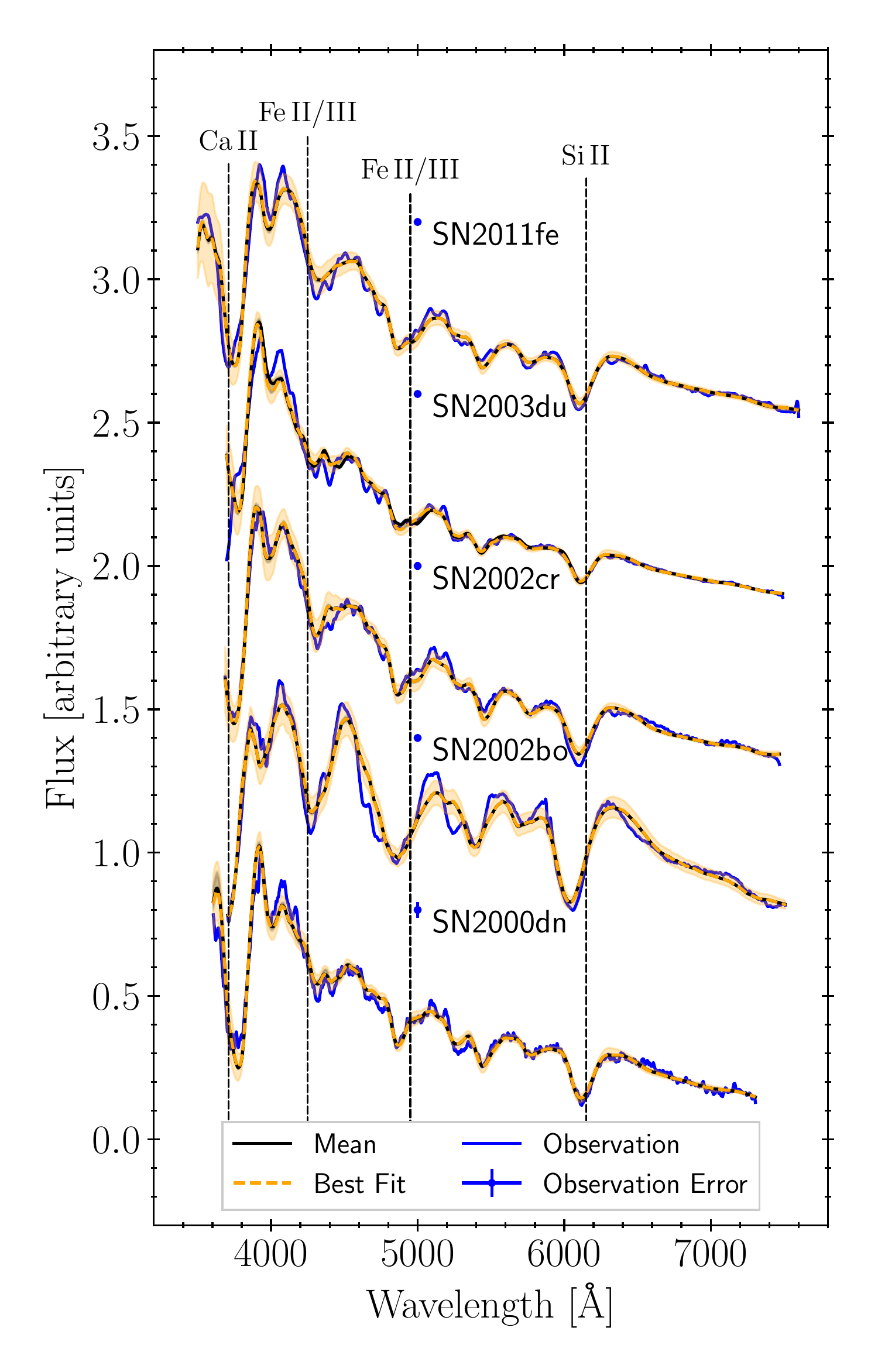} 
  
  \caption{Same as Figure~\ref{fig:Spectra} but for a selection of five normal SNe~Ia for comparison.}
  \label{fig:Spectra2}
\end{figure}

The posterior probability distribution of spectra for the five 1991T-like SNe~Ia in our samples along with their maximum likelihood estimate and total uncertainty is presented in Figure~\ref{fig:Spectra}. For comparison, a selection of five of the normal SNe~Ia from our sampled are shown in Figure~\ref{fig:Spectra2}. Our fits accurately reproduce major line features that distinguish 1991T-like SNe~Ia from the normal SNe~Ia population. Specifically, our models generate the high-velocity Fe~III features around $\num{4250}$\,\textrm{\AA} and $\num{4950}$\,\textrm{\AA} as well as the Si~II feature near $\num{6150}$\,\textrm{\AA}.

\subsection{Ejecta Properties}
\label{sec:EjectaProperties}

The peculiar nature of early-time 1991T-like spectra has been well identified, but their origin remains unclear. 1991T-like spectra show the presence of high-velocity Fe~III emission and lack the strong characteristic Si~II and Ca~H\&K absorption commonly seen in Branch-normal SNe~Ia \citep[see e.g.][]{Filippenko1997}. After maximum light, 1991T-like spectra begin to behave similarly to normal Type Ia spectra, with Si~II features reappearing in the spectra \citep[see e.g.][]{2017hsn..book..317T}. There have been two suggested causes behind the lack of singly-ionized IME absorption at early times. Namely, a lack of total IME production and higher ionization states of IMEs produced in the ejecta \citep[e.g.][]{1992ApJ...397..304J,1992ApJ...387L..33R,10.1093/mnras/stu1777} .

\begin{figure}

  \centering
  \includegraphics[width=0.5\textwidth]{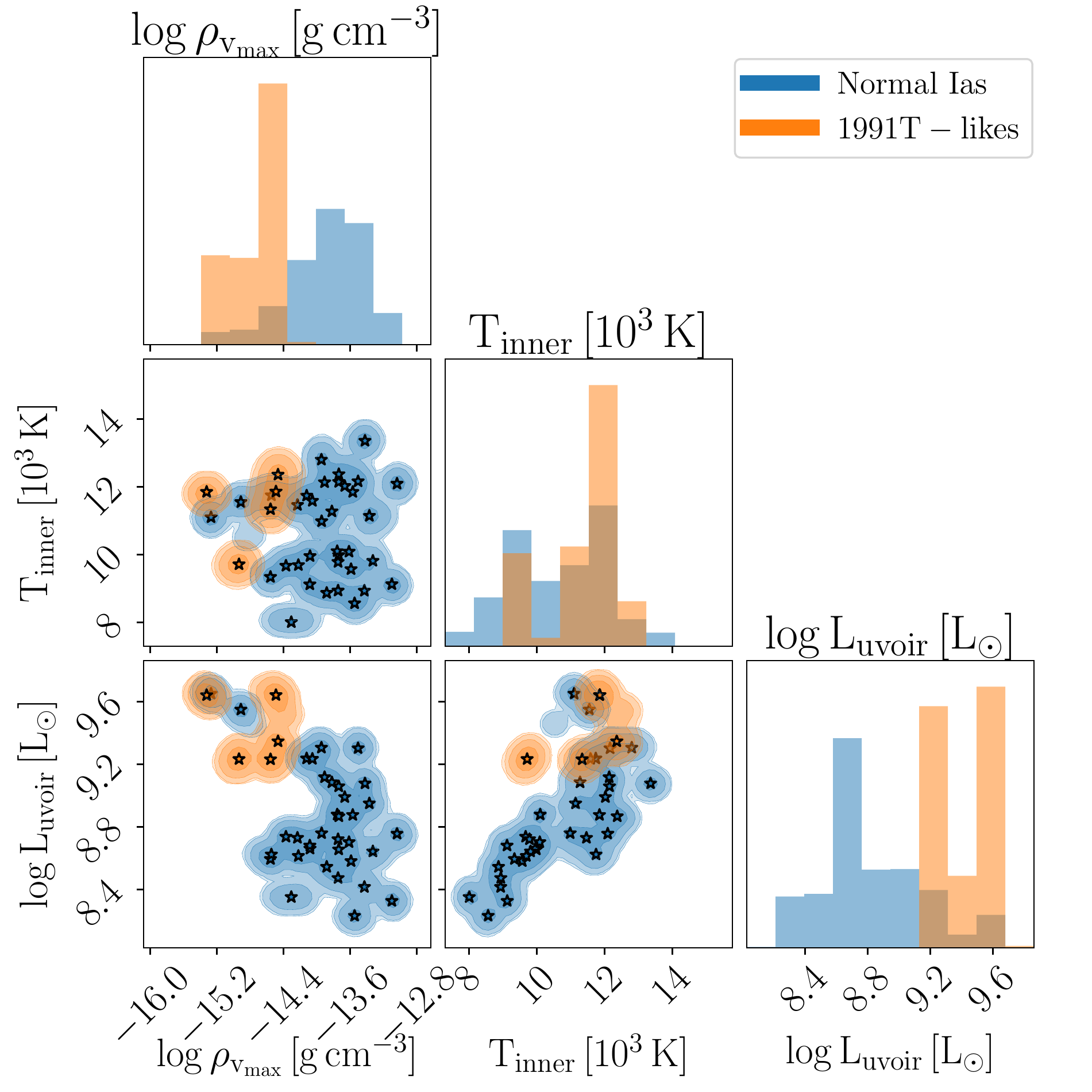}
  \caption{Posterior contours of SNe~Ia probed in this study. Normal SNe~Ia are shown with blue contours and 1991T-like SNe~Ia are shown in orange. The contours cover 68\%, 95\%, and 99.5\% quantiles. The posterior means for each object are shown as stars. The plots show the joint distributions between ejecta density at $v=v_\mathrm{max}$, inner boundary temperature $T_\mathrm{inner}$, and integrated UVOIR luminosity from the model spectrum. 
  While 1991T-like SNe~Ia are generally brighter than the Normal SNe~Ia population, the increase in brightness does not seem to be driven by substantially higher photospheric temperatures. The lower ejecta density in the region of highest intermediate mass element abundance shows that higher ionization fractions in 1991T-like SNe~Ia are influenced by the lower electron density.}
  \label{fig:EjectaStructure}

\end{figure}

We find a variety of parameters that indicate the differences between 1991T-like and Normal SNe~Ia. The distribution of inner boundary temperatures for 1991T-like SNe~Ia does not substantially differ from those of normal SNe~Ia (Figure~\ref{fig:EjectaStructure}) indicating that high-ionization states of IMEs, in particular Silicon, are not due to a difference in temperature of the ejecta alone. This leads us then to investigate two other possible causes for the lack of Si~II formation in the photospheric phase: a decrease in the electron density at the primary location of IME composition or a decrease in the total mass of IMEs contributing to the line features seen in the ejecta.

The material below the photosphere, parameterized through the inner boundary velocity, does not contribute to features in the resultant spectra. Therefore, constraints of physical properties of the ejecta must rely strictly upon material above the inner boundary photosphere. We determine the total mass of each contributing element above the photosphere by integrating Equations~\ref{eq:MIME},~\ref{eq:MIGE},~and~\ref{eq:MUBE} with their lower bounds set to the inner boundary velocity, $v_\mathrm{inner}$. We compute the mass fraction of each element as the integrated mass of each element above the photosphere divided by the total mass above the photosphere. The mass fraction offers a direct probe of the nucleosynthetic products that are visible in the photospheric phase and which can be directly compared to hydrodynamic models without a need to convert abundance fractions into total masses.

\begin{figure}
  \centering
  \includegraphics[width=0.5\textwidth]{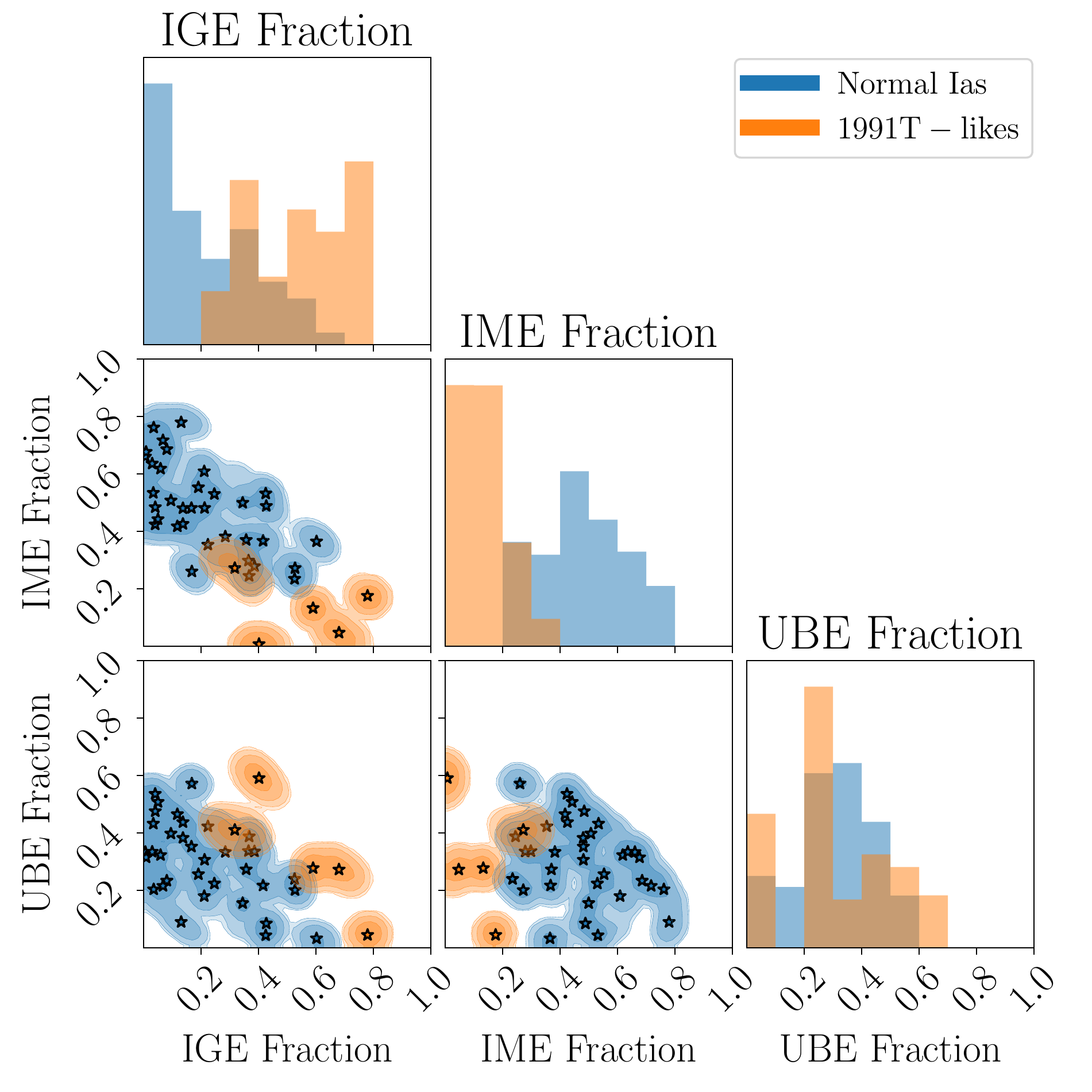}
  \caption{Posterior contours of SNe~Ia probed in this study. Normal SNe~Ia are shown with blue contours and 1991T-like SNe~Ia are shown in orange. The contours cover 68\%, 95\%, and 99.5\% quantiles. The posterior means for each object are shown as stars. Each plot shows the marginal distribution of mass fractions of the various ejecta compositions above the inner boundary velocity by integrating equations \ref{eq:MIGE}, \ref{eq:MIME}, and \ref{eq:MUBE}. It can clearly be seen that 1991T-like SNe~Ia lie on the edge of IME mass fraction distribution describing normal SNe~Ia.}
  \label{fig:ElemMassFraction}
\end{figure}

Figure~\ref{fig:ElemMassFraction} shows the posterior probability distributions of the IME fractions from 1991T-like SNe~Ia demonstrating a clear deficit compared to that of normal SNe~Ia coupled with a small increase of IGEs as a fraction of the total ejecta. The marginal distribution of the fraction of unburned elements does not demonstrate a discernible difference between 1991T-likes and Normal SNe~Ia, though the joint distribution between IGEs and unburned elements shows an interesting correlation in 1991T-likes in which the fraction of unburned elements in the ejecta is slightly higher for 1991T-like SNe~Ia compared to Normal SNe~Ia given the same iron-group element fraction. The consistent lack of IME mass fractions changing with respect to UBE fractions along with the correlation between UBE and IME fractions in 1991T-like SNe~Ia implies a rapid and consistent drop-off in the rate of production of nucleosynthetic products with respect to depth into the explosion.

While many 1991T-like SNe~Ia show generally lower mass fractions of IMEs compared to the normal Ia population, there are cases of overlap (see Figures~\ref{fig:EjectaStructure},~\ref{fig:ElemMassFraction}) where low mass fractions alone are not enough to explain the observed lack of IME features, such as the Si~II $\num{6150}$\,\textrm{\AA} doublet, in the resulting spectra. Additionally, we note that the 1991T-like SNe~Ia population has generally lower ejecta densities at the location of the peak of the fractional abundance of IMEs in our model implying a lower electron density and therefore a higher ionization state. The combination of low IME mass fraction and higher ionization states leads to a dual effect where the observed properties of 1991T-like SNe~Ia in comparison to the normal Ia population is not due to a single underlying mechanism, but a combination of different physical processes which result in similar looking spectra observationally.

\begin{figure}
  \centering
  \includegraphics[width=0.5\textwidth]{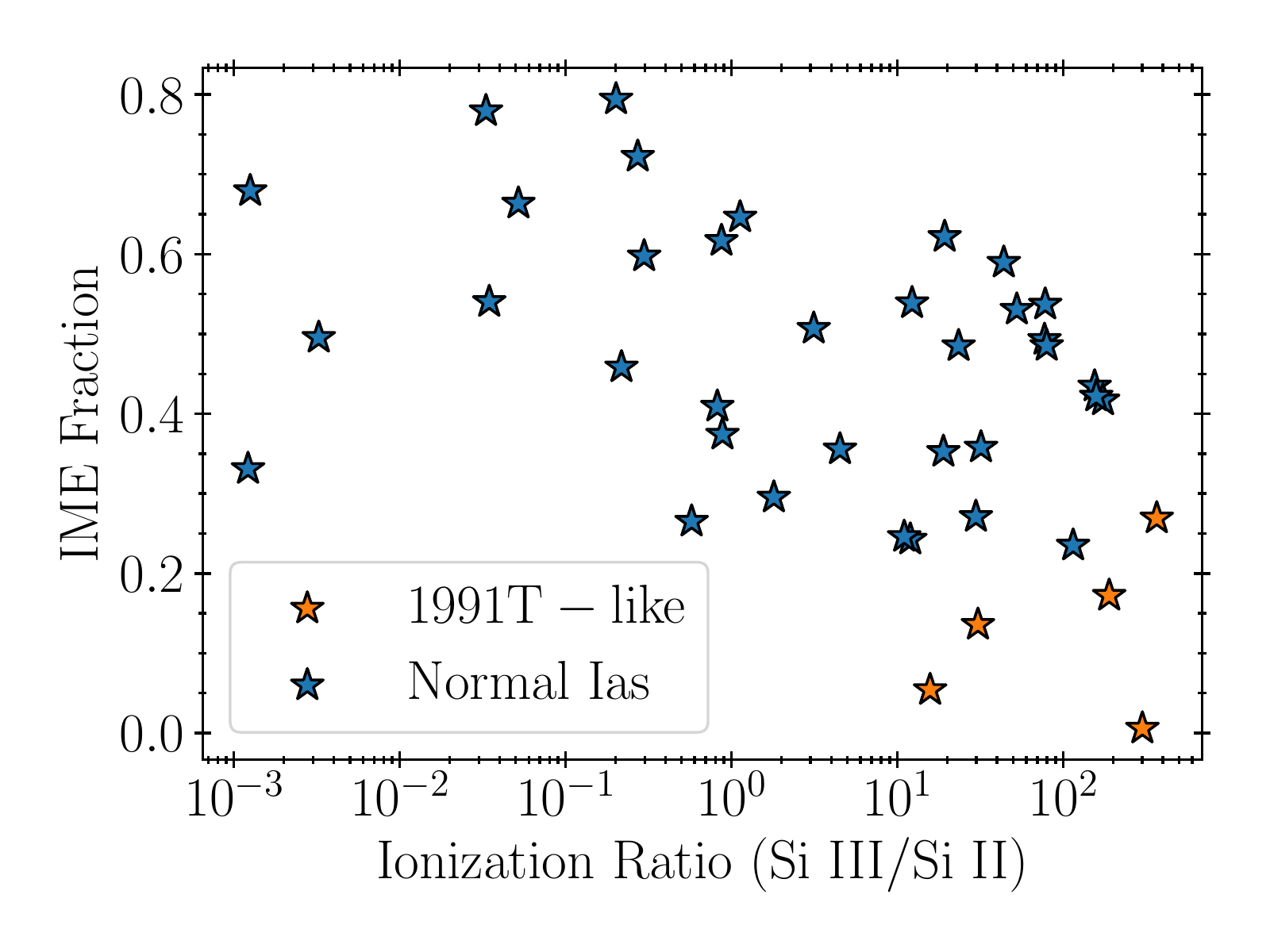}
  \caption{Ratio of Si~III to Si~II ion density at the shell encompassing $v_{max}$ versus the fraction of intermediate-mass elements in the ejecta above the inner boundary. Samples are taken from the maximum likelihood parameters of each SNe~Ia in our sample run through \textsc{tardis} to calculate the properties of the radiation field. Normal SNe~Ia are shown in blue and 1991T-like SNe~Ia are shown in orange. A clear turnover between normal SNe~Ia and 1991T-like SNe~Ia is evident in the regions of low intermediate mass element fraction and high ionization state of silicon. The combination of the lack of material coupled with high ionization states creates a boundary between the spectral types.}
  \label{fig:IonizationStateScatter}
\end{figure}

We selected the maximum likelihood sample for each spectrum and simulated the ejecta radiation field using \textsc{tardis}. The ionization fraction of Si~III to Si~II was determined at the shell containing the velocity $v_{max}$ providing a look into the ionization state of the plasma at the location with the maximum abundance of intermediate-mass elements. 1991T-like supernovae were found to have overall higher ionization fractions than the vast majority of normal SNe~Ia, though some overlap was found within the normal SNe~Ia population (Figure~\ref{fig:IonizationStateScatter}). The normal SNe~Ia with comparable ionization fractions to 1991T-like SNe~Ia all had a higher mass fraction of IMEs than 1991T-like SNe~Ia at the same ionization fraction. Additionally, 1991T-like SNe~Ia with lower ionization fractions among the 1991T-like population also had a lower mass fraction of intermediate-mass elements within their ejecta. The suppressed Si~II absorption features observed in 1991T-like SNe~Ia, therefore, is a result of a combination of low IME fraction and high ionization state, creating a region of space where there is a turnover in the classification between the spectral types.

\section{Conclusion}
\label{sec:Conclusion}

We have performed single-epoch ejecta reconstructions for 35 normal SNe~Ia and five 1991T-like SNe~Ia. Elemental abundance distributions and their ionization fractions have provided a picture linking the internal properties of the ejecta to the observational properties of their spectra. Comparison between the ejecta properties between the two populations provides insight into the relationship between normal SNe~Ia and 1991T-like SNe~Ia.

We find 1991T-like SNe~Ia both under-produce IMEs relative to the normal SNe~Ia population and these IMEs exist in higher ionization states than the IMEs in the normal SNe~Ia population. The cause of the higher ionization fractions is primarily driven by a lower overall electron density in the ejecta. The lower overall electron density may be a result of a relative overabundance of IGEs relative to the abundance of IMEs in the ejecta of 1991T-like SNe~Ia resulting in an ejecta composition dominated by high-neutron number elements, while normal SNe~Ia with depleted IMEs may have the remainder of the ejecta filled with unburned Carbon and Oxygen. 

Neither the low abundance fraction of IMEs nor the high ionization states of IMEs alone are enough to explain the peculiar properties of 1991T-like SNe~Ia; instead, a combination of the two effects drives their unique spectral signatures at early times. We have found 1991T-like SNe~Ia that contain a similar IME fraction to some of the normal SNe~Ia in our sample, but these 1991T-like SNe~Ia have a higher overall IME ionization than a normal SNe~Ia at a similar IME mass fraction. Conversely, we have found 1991T-like SNe~Ia with similar IME ionization fractions to the normal SNe~Ia but these objects have a lower mass fraction of IMEs than the normal SNe~Ia given their ionization state.

Our findings suggest that normal SNe~Ia and 1991T-like SNe~Ia might arise from a similar population or progenitor system. The observational spectral properties that traditionally separate the two groups result from a sharp change in the amplitude of spectral features corresponding to IMEs over small changes in both composition and ionization state. This results in small deviations in ejecta composition leading to a sharp contrast in observed spectral features.

\section{Acknowledgments}
\label{sec:Acknowledgments}

This work was supported in part through computational resources and services provided by the Institute for Cyber-Enabled Research at Michigan State University.\\
This work made use of the Heidelberg Supernova Model Archive (\textsc{HESMA}), \url{https://hesma.h-its.org}\\
This research made use of \textsc{tardis}, a community-developed software package for spectral
synthesis in supernovae \citep{TARDIS2014, kerzendorf_wolfgang_2021_4995779}. The
development of \textsc{tardis} received support from the Google Summer of Code initiative,
from ESA's Summer of Code in Space program, and from NumFOCUS's Small Development Grant. \textsc{tardis} makes extensive use of Astropy\footnote{\url{https://www.astropy.org}} \citep{astropy:2013,astropy:2018} \\
C.V. was supported for this work by the Excellence Cluster ORIGINS, which is funded by the Deutsche Forschungsgemeinschaft (DFG, German Research Foundation) under Germany's Excellence Strategy -- EXC-2094 -- 390783311.\\

\software{Matplotlib\footnote{\url{https://matplotlib.org}} \citep{matplotlib}, Numba\footnote{\url{https://numba.pydata.org}} \citep{numba}, NumPy\footnote{\url{https://numpy.org}} \citep{numpy}, pandas\footnote{\url{https://pandas.pydata.org}} \citep{pandas}, scikit-learn\footnote{\url{https://scikit-learn.org}} \citep{sklearn}, SciPy\footnote{\url{https://www.scipy.org/}} \citep{scipy}, \textsc{tardis}\footnote{\url{https://tardis-sn.github.io/tardis}} \citep{TARDIS2014},
Pytorch\footnote{\url{https://pytorch.org/}} \citep{pytorch}, and UltraNest\footnote{\url{https://johannesbuchner.github.io/UltraNest}} \citep{Buchner2014,Buchner2019}}

\bibliographystyle{aasjournal}
\bibliography{sources}

\section{Contributor Roles}
\begin{itemize}
\item Conceptualization:
John O'Brien, Wolfgang Kerzendorf, R\"udiger Pakmor
\item Data curation: John O'Brien
\item Formal Analysis:
John O'Brien
\item Funding acquisition:
Wolfgang Kerzendorf
\item Investigation: John O'Brien
\item Methodology:
John O'Brien, Wolfgang Kerzendorf, Nutan Chen, Patrick van der Smagt, Johannes Buchner, Christian Vogl
\item Project administration:
Wolfgang Kerzendorf
\item Resources: Institute for Cyber-Enabled Research at Michigan State University
\item Software:
John O'Brien,
Nutan Chen,
Wolfgang Kerzendorf,
Johannes Buchner,
Andrew Fullard,
Christian Vogl,
\textsc{tardis} Collaboration
\item Supervision:
Wolfgang Kerzendorf
\item Validation: 
John O'Brien
\item Visualization: 
John O'Brien,
Jaladh Singhal
\item Writing – original draft:
John O'Brien,
Wolfgang Kerzendorf
\item Writing – review \& editing:
John O'Brien,
Wolfgang Kerzendorf,
Andrew Fullard,
Johannes Buchner,
Christian Vogl,
Patrick van der Smagt,
Nutan Chen,
R\"udiger Pakmor,
Marc Williamson
\end{itemize}

\appendix

\section{Data Products} \label{Appendix:A}

\textsc{tardis} configuration information, emulator weights and training data, and example scripts can be found at the following location can be found at the following link: \href{https://doi.org/10.5281/zenodo.7818303}{10.5281/zenodo.7818303}

\end{document}